\documentclass[12pt,english,reprint,aps,prb,superscriptaddress]{revtex4-1}
\usepackage[T1]{fontenc}
\usepackage[utf8]{inputenc}
\usepackage{amstext}
\usepackage{amssymb}
\usepackage{graphicx}

\makeatletter
\usepackage{babel}
\usepackage[unicode=true]{hyperref}
\usepackage{newtxtext}
\usepackage[slantedGreek,varg,varbb]{newtxmath}
\usepackage{mathtools}
\usepackage{siunitx}
\usepackage[normalem]{ulem}
\usepackage{xcolor}
\providecolor{added}{rgb}{0,0,1}
\providecolor{deleted}{rgb}{1,0,0}

\makeatother

\usepackage{babel}
\begin{document}

\title{Conductance resonances and crossing of the edge channels in the quantum Hall ferromagnet state of Cd(Mn)Te microstructures}

\author{E.~Bobko}
\email{baranewa@ur.edu.pl}

\affiliation{Faculty of Mathematics and Natural Sciences, Rzeszów University,
Aleja Rejtana 16A, PL-35-959 Rzeszów, Poland}

\affiliation{Institute of Physics, Polish Academy of Sciences, Aleja Lotników
32/46, PL-02-668 Warszawa, Poland}

\author{D.~Płoch}

\affiliation{Faculty of Mathematics and Natural Sciences, Rzeszów University,
Aleja Rejtana 16A, PL-35-959 Rzeszów, Poland}

\author{D.~Śnieżek}

\affiliation{Institute of Physics, Polish Academy of Sciences, Aleja Lotników
32/46, PL-02-668 Warszawa, Poland}

\author{M.~Majewicz}

\affiliation{Institute of Physics, Polish Academy of Sciences, Aleja Lotników
32/46, PL-02-668 Warszawa, Poland}

\author{M.~Fołtyn}

\affiliation{Institute of Physics, Polish Academy of Sciences, Aleja Lotników
32/46, PL-02-668 Warszawa, Poland}

\author{T.~Wojciechowski}

\affiliation{Institute of Physics, Polish Academy of Sciences, Aleja Lotników
32/46, PL-02-668 Warszawa, Poland}

\affiliation{International Research Centre MagTop, Aleja Lotników 32/46, PL-02-668
Warszawa, Poland}

\author{S.~Chusnutdinow}

\affiliation{Institute of Physics, Polish Academy of Sciences, Aleja Lotników
32/46, PL-02-668 Warszawa, Poland}

\author{T.~Wojtowicz}

\affiliation{Institute of Physics, Polish Academy of Sciences, Aleja Lotników
32/46, PL-02-668 Warszawa, Poland}

\affiliation{International Research Centre MagTop, Aleja Lotników 32/46, PL-02-668
Warszawa, Poland}

\author{J.~Wróbel}

\affiliation{Faculty of Mathematics and Natural Sciences, Rzeszów University,
Aleja Rejtana 16A, PL-35-959 Rzeszów, Poland}

\affiliation{Institute of Physics, Polish Academy of Sciences, Aleja Lotników
32/46, PL-02-668 Warszawa, Poland}

\begin{abstract}
In this paper we report on the observation of very high and narrow magnetoconductance peaks which we attribute to the transition to quantum Hall ferromagnet (QHFM) state occurring at the edges of the sample. We show that the expected spatial degeneracy of chiral edge channels is spontaneously lifted in agreement with theoretical studies performed within the Hartree-Fock approximation. We indicate also that separated edge currents which flow in parallel, may nevertheless cross at certain points, giving rise to the formation of \textit{topological defects} or one-dimensional magnetic domains. Furthermore, we find that such \emph{local crossing} of chiral channels can be induced on demand by coupling spin states of a Landau level to different current terminals and applying a DC source-drain voltage.
\end{abstract}
\maketitle

\section{Introduction}

The Pauli principle does not forbid the crossing of quantum Hall edge channels formed at the boundary of a 2D electron gas (2DEG), provided that they are labeled by different quantum numbers. For spin degenerate states, however, a short-ranged exchange interaction enhances Zeeman energy and leads to the reconstruction of spin-split currents that become spatially separated \cite{Dempsey1993}. The distance between such channels is dictated by a long-rang Hartree term and by the strength of the confining potential \cite{Chklovskii1992,Meir1994}. Therefore, the spatial separation cannot be continuously reduced to zero and overlap  is prohibited in the vicinity of smooth edges. More recently this analysis was extended to the helical states in 2D topological insulators \cite{Wang2017}. The reconstruction of edge currents has been already observed in quantum Hall effect experiments for both integer \cite{Klein1995} and fractional \cite{Sabo2017} filling factors.

In order to circumvent the electrostatic repulsion, Rijkels and Bauer \cite{Rijkels1994} proposed to drive the spin states of a Landau level into a non-equilibrium configuration. They studied the case of a chemical potential difference between states  in the Hartree-Fock approximation and showed that it leads to a \emph{local reversal} of edge channel positions or, in other words, to a soliton-like crossing of boundary states. These authors showed that such topological defects are in fact point-like Bloch walls between 1D magnetic domains which can be used for spin filtering.

From today's perspective the properties of these defects may be even more interesting than anticipated by Rijkels and Bauer. Edge currents formed in the QHE regime are nowadays widely studied as prototypes of a \emph{chiral} 1D Luttinger liquid, which is not time-reversal invariant and has the property of spin-charge separation. This was recently confirmed in experiments \cite{Bocquillon2013a}. With the advent of reliable mesoscopic one-electron sources coupled coherently to edge channels\cite{Feve2007}, the intersection of chiral wave-guides will definitely find numerous applications in the field of electron quantum optics, especially if it can be created on-demand.

The experimental signatures of point-like defects in the edge-state structure have already been reported for the special Corbino geometry \cite{Deviatov2005}, however in that case, the intersecting currents resided in separate quantum wells of a bi-layer electron system. To our knowledge, the crossing of spin-split chiral states that flow at the boundary of a standard device made of a single quantum well, has not been observed experimentally. In this paper, confirming the prediction of Rijkels and Bauer, we show that edge currents which are not in equilibrium cross at a certain point, when a single layer 2DEG system is tuned \emph{locally} to the quantum Hall ferromagnet (QHFM) state.

\section{Diluted magnetic quantum well}

\begin{figure}
	\begin{centering}
		\includegraphics[scale=0.8]{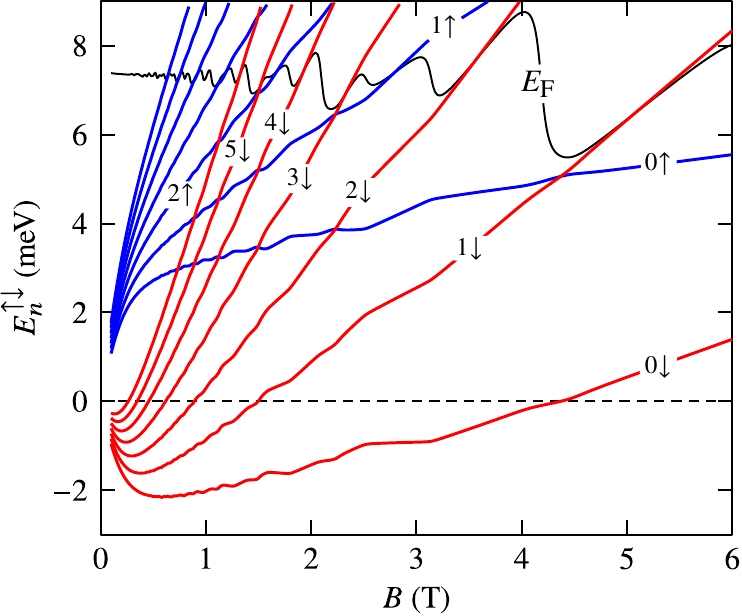} 
		\par\end{centering}
	\caption{Energies of spin-split Landau levels $E_{n}^{\uparrow\downarrow}$ vs. magnetic field $B$ (up to $n=6$) calculated self-consistently using a modified Zeeman term taken from Eq.~(\ref{eq:deltas}). Spin-up $(n\uparrow)$ and spin-down $(n\downarrow)$ ladders are indicated with blue and red colors, respectively; The black solid line shows the position of the Fermi energy $E_{\mathrm{F}}$. The chemical potential and spin polarization were calculated at temperature $T=0.050\;\si{\kelvin}$ assuming $\Gamma=0.1\;\si{\milli\electronvolt}$ as the Gaussian broadening of the density of states \cite{Kunc2015}. \label{fig:LLcalc}}
\end{figure}

Modulation doped Cd$_{1-x}$Mn$_{x}$Te semiconductor quantum wells (QWs) possess a very large electronic Lande-factor, which depends on the magnetization of the Mn ions via the \emph{s-d} exchange interaction \cite{Furdyna1988}. In consequence, Landau levels (LLs) with spin up and spin down energies $E_{n}^{\uparrow}$ and $E_{m}^{\downarrow}$ cross at certain values of the magnetic field, as shown in Fig.~\ref{fig:LLcalc}. The most prominent transition to the QHFM state occurs when the condition $E_{0}^{\uparrow}=E_{1}^{\downarrow}$ is fulfilled for energy which coincides with the chemical potential.

If carrier density corresponds to a filling factor of $\nu=2$, electrons are tuned by the magnetic field from the polarized (ferromagnetic) to an unpolarized spin configuration. As a result, a phase transition driven by electron-electron (\emph{e-e}) exchange interaction, occurs at some critical field $B_{\mathrm{c}}$ \cite{Giuliani1985}. The new QHFM phase is characterized by a non-uniform texture of spin polarization because the electron liquid is spontaneously divided into 2D ferromagnetic and antiferromagnetic domains \cite{Jungwirth2001}. Experimentally, the crossing point of the LLs manifests itself as an additional and rather narrow resistance $\rho_{xx}$ peak (or a conductance $\sigma_{xx}$ peak). This was indeed observed for diluted magnetic Cd(Mn)Te quantum wells by Jaroszyński \emph{et.al.} \cite{Jaroszynski2002}. Moreover, the shape of the $\rho_{xx}(B)$ curve around $B_{\mathrm{c}}$ was dependent on the sweep direction of magnetic field. It is widely accepted, that such \emph{hysteretic} behavior proves the existence of magnetic domains. Electron transport then occurs through a conducting network of randomly connected domain walls \cite{Jungwirth2001,Jaroszynski2002}.

The quantum Hall ferromagnet state for integer filling factors $\nu$ was observed for the first time in non-magnetic semiconductors \cite{Koch1993} and since then it has been extensively studied \cite{Chokomakoua2004}, also for the fractional quantum Hall effect regime \cite{Smet2001}. Recent renewed interest in the QHFM transition, in particular for diluted magnetic materials, stems from the observation that domain walls that form a percolation path are in fact \emph{helical} 1D channels. These have potential application to generate non-Abelian excitations \cite{Kazakov2017}.

\section{Interacting chiral channels}

When describing the QHFM ground state which is formed at integer $\nu$ in macroscopic samples, attention is usually focused on the bulk properties of the magnetic phases, and it is emphasized that spin splitting is enhanced when level crossings occur in the vicinity of chemical potential. Here, we would like to examine the consequences of LL crossings which take place \emph{below} Fermi energy in mesoscopic micro-structures, using the language of Quantum Hall edge currents. In this picture, the bulk filling factor decreases in a step-wise manner towards the edge of a device and reaches zero at the physical boundary of the 2DEG system \cite{Chklovskii1992}. As a consequence, for a structure made of Cd$_{1-x}$Mn$_{x}$Te QW, we expect a series of quasi-one-dimensional QHFM transitions to occur along the edges. They may be detectable by non-local measurements of the magnetoconductance.

\begin{figure}
	\begin{centering}
		\includegraphics[scale=0.8]{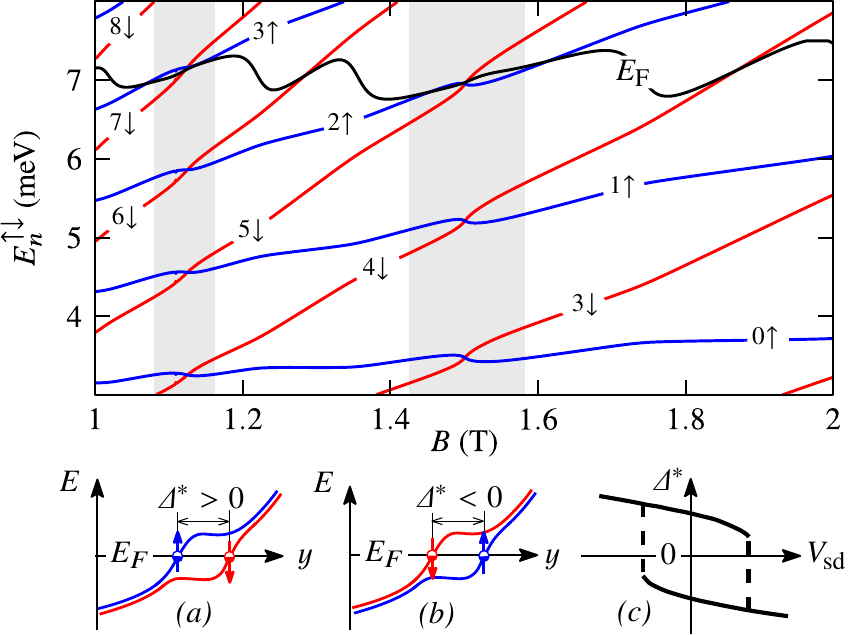} 
		\par\end{centering}
	\caption{Spin-up $E_{n}^{\uparrow}$ (blue), spin-down $E_{n}^{\downarrow}$ (red) Landau levels and Fermi energy $E_{\mathrm{F}}$ calculated for magnetic field range \SIrange{1}{2}{\tesla}. Below, edge current picture of LLs crossings occurring at $B=B_{\mathrm{c}}$ within the shaded bands, as indicated above. (a) Position dependent energies of $(n\uparrow)$ and $(m\downarrow)$ LLs for $B\lesssim B_{\mathrm{c}}$, $y$ is the spatial coordinate, $\Delta^{*}=y_{\downarrow}-y_{\uparrow}$. (b) The same for $B\gtrsim B_{\mathrm{c}}$. (c) $\Delta^{*}$ \emph{vs} source-drain voltage $V_{\mathrm{sd}}$ (adopted from \cite{Rijkels1994}). \label{fig:edge-1}}
\end{figure}

As a first step towards the verification of these ideas, we calculated $E_{n}^{\uparrow\downarrow}(B)$ with the addition of a term that accounts for the electron-electron interactions. Such a correction is necessary because correlation effects in Cd$_{1-x}$Mn$_{x}$Te are stronger than e.g. in the well known case of GaAs. Indeed, it has been shown experimentally for a 2D electron gas confined in a CdTe QW that a considerable enhancement of the spin gap occurs for fully occupied LLs below the Fermi energy $E_{\mathrm{F}}$. Moreover, it has been argued that such enhancement is driven by the \emph{total} spin polarization and results in a rigid shift of the LL ladders with opposite spins \cite{Kunc2010}. Using this assumption, magnetoluminescence and magnetotransport measurements can be successfully described using the following simple formula for a total spin splitting:
\begin{equation}
\Delta_{s}=|g^{*}|\mu_{\mathrm{B}}B+\delta_{0}\thinspace\sqrt[4]{B^{2}+B_{0}^{2}}\thinspace\mathcal{P}.\label{eq:deltas}
\end{equation}
It is common for all Landau levels. Here $\mathcal{P}=(n_{\downarrow}-n_{\uparrow})/(n_{\downarrow}+n_{\uparrow})$ is the spin polarization and $g^{*}$ is an effective Lande factor of the electrons. Phenomenological constants $\delta_{0}\sqrt{B_{0}}=2.1\ \si{\milli\electronvolt}$ and $B_{0}=3.7\ \si{\tesla}$ have been determined as fitting parameters \cite{Kunc2010}.

Following \cite{Kunc2015}, we have assumed that those findings apply also for Cd$_{1-x}$Mn$_{x}$Te quantum wells. Accordingly, we have calculated $E_{n}^{\uparrow\downarrow}$ in a self-consistent way, taking into account the fact that $\Delta_{s}$ depends on polarization $\mathcal{P}$, which in turn depends on the total spin splitting, as it follows from Eq.~\ref{eq:deltas}. For the effective $g$-factor in diluted magnetic materials we used a standard formula \cite{Furdyna1988}. We assumed that electron density $n_{\text{2D}}$ is fixed (see below) and the Mn content $x$ was treated as a fitting parameter. We obtained $x=0.0096$. Results for $n_{\text{2D}}=\num{3.1E11}\thinspace\si{\per\centi\meter\squared}$ and temperature $T=0.05\thinspace\si{\kelvin}$ are shown in Figs~\ref{fig:LLcalc} and \ref{fig:edge-1}.

In the language of non-interacting edge currents, such crossing of energy levels corresponds to the situation when two channels, belonging to distinct Landau states and carrying electrons with opposite spins, overlap in real space. This picture is, however, too simplified since correlation effects lift the spatial degeneracy of chiral channels, as explained before. Such \emph{reconstruction} of spin-split edge channels is schematically illustrated in Fig~\ref{fig:edge-1}a,b for a quantum structure made from a Cd$_{1-x}$Mn$_{x}$Te material. Note that the distance $\Delta^{*}$ between chiral states changes sign as a function of perpendicular magnetic field. Therefore, during such reordering, channels must overlap at critical field, in spite of the reconstruction effect. In fact, Rijkes and Bauer \cite{Rijkels1994} showed that $\Delta^{*}$ demonstrates jumps and \emph{bistable} behavior dependent on source-drain voltage $V_{\mathrm{sd}}$ in the non-linear transport regime (see Fig.~\ref{fig:edge-1}c).

\section{Experiment and results}

\begin{figure}
	\begin{centering}
		\includegraphics[scale=0.9]{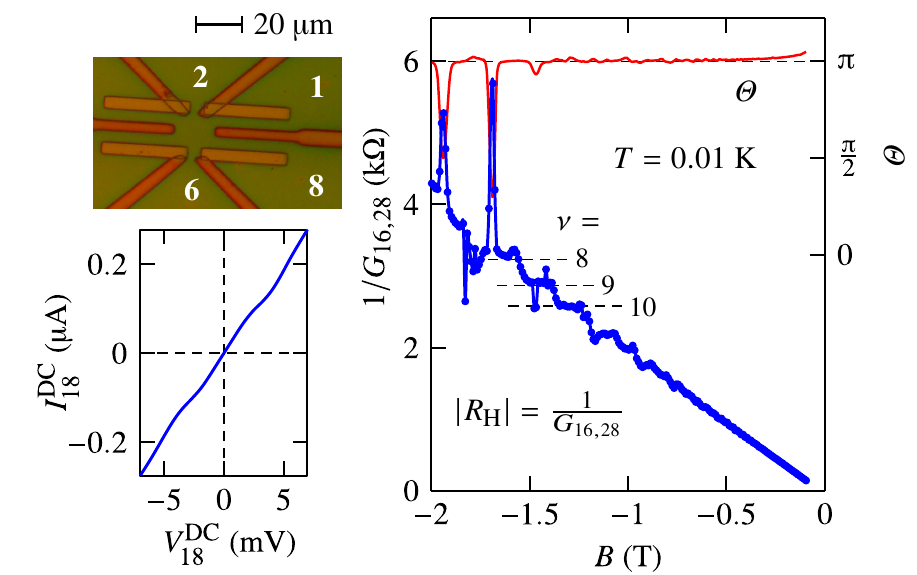} 
		\par\end{centering}
	\caption{Hall resistance $R_{\mathrm{H}}$ and phase of Hall voltage $\Theta$ vs magnetic field for $B<0$. An image of the sample with numbered contact areas and separating grooves is shown on the left. Pads (1) and (6) were used as current contacts, (2) and (8) as voltage probes. Non-labeled contacts were not used. Below, the $I$-$V$ characteristic of (1-8) terminal pair is also shown, data were recorded at $B=1.508$~T.} \label{fig:R1628+sample}
\end{figure}

To experimentally test these theoretical predictions we have performed low-temperature magneto-transport measurements on a \textit{quasi-}ballistic micro-structure produced from a $30$-nm-wide Cd$_{1-x}$Mn$_{x}$Te QW with a manganese content of $x\approx0.01$. The 2DEG structure was grown by molecular beam epitaxy (MBE) with Cd$_{1-u}$Mg$_{u}$Te barriers ($u=0.31$), and modulation doped with iodine donors which were introduced into the top barrier at a distance of $300\;\text{Å}$ from the QW in the growth direction. The 6-terminal microstructure shown in the inset of Fig.~\ref{fig:R1628+sample} was patterned on this substrate using \textit{e}-beam lithography and deep-etching techniques. The sample was formed into a \textsf{H}-shaped device with an overall size of $15\ \si{\micro\metre}$, connected to the external contacts by 4 multimode wires of the same length ($\approx30\;\si{\micro\meter}$) and of geometrical width $W_{\mathrm{geo}}$, which was slightly different for adjacent channels. In particular, $W_{\mathrm{geo}}=5\pm0.1\;\si{\micro\meter}$ and $6\pm0.1\;\si{\micro\meter}$ for wires 1 and 8, respectively (see Fig.~\ref{fig:R1628+sample}). Additionally, two much shorter and slightly narrower constrictions of $W_{\mathrm{geo}}=3\pm0.1\;\si{\micro\meter}$ were connected to the inner part of the device (terminals 2 and 6). 

\begin{figure}
	\begin{centering}
		\includegraphics[scale=0.9]{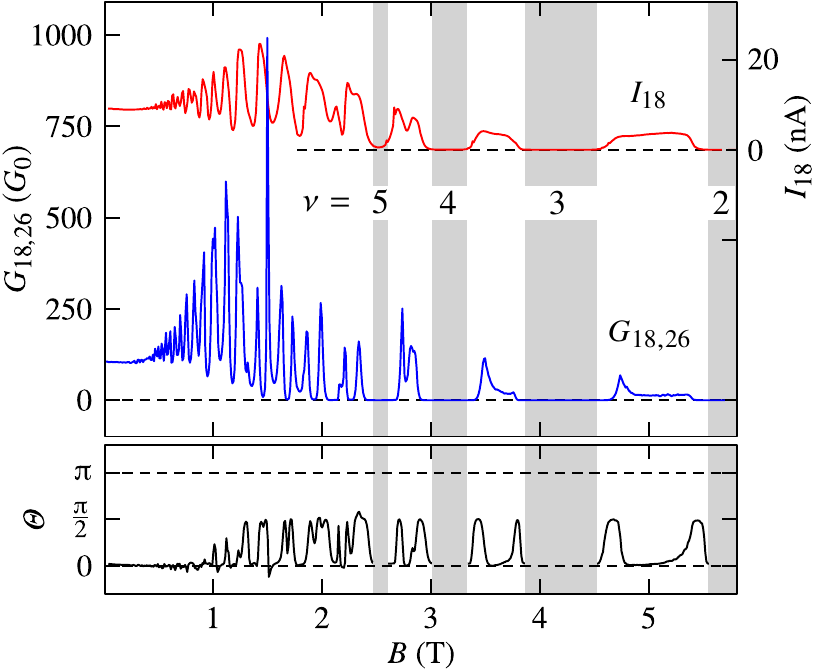} 
		\par\end{centering}
	\caption{$G_{18,26}$ (in $G_{0}=2e^{2}/h$ units) and current $I_{18}$ at a base temperature $T=0.01\;\si{\kelvin}$ vs increasing magnetic field $B$. The relative phase $\Theta$ of the $V_{\mathrm{AC}}$ and $V_{26}$ signals vs magnetic field $B$ is shown below. Shaded bands represent the magnetic field ranges, where $I_{18} \rightarrow 0$, corresponding filling factors $\nu$ are also indicated.}
	\label{fig:G1826-G-Theta}
\end{figure}

Special attention was devoted to the fabrication of macroscopic contacts to our device. Typically, electrical connections to Cd(Mn)Te quantum wells are obtained with macroscopic indium droplets, which are burnt-in or soldered locally at the edges of a sample. Such procedures are not fully reproducible and clearly, they are not compatible with e-beam lithography. Nevertheless, they are widely adopted, because the  global annealing at temperatures above $\approx 200 \si{\celsius}$ is not recommended for II-VI materials. In this work, we prepared the soldering pads lithographically, using mercury telluride (HgTe) zero-gap semiconductor as a low-resistivity contact material \cite{Jaroszynski1983}. The lowest resistances were obtained when HgTe was thermally evaporated on substrates cooled with liquid nitrogen, in this case the polycrystalline contact material was almost perfectly stoichiometric. After evaporation, the HgTe surface was covered with Cr/Au metal layer and heated locally during the subsequent process of soldering of indium droplets. Contact pads extended to the edges of the device and local annealing occurred far away from the central region of the microstructure. Contacts possessed linear $I$-$V$ characteristics with the zero-bias resistances $R_{\mathrm{k}}=5\;\si{\kilo\ohm}$ to $13\;\si{\kilo\ohm}$ (for $B\gtrsim0.02\;\si{\tesla}$) and were highly reproducible during the  thermal cycling. An example of the $I$-$V$ curve, for two contacts connected in series, is shown in Fig.~\ref{fig:R1628+sample}.

We studied the local and non-local differential magnetoconductances $G_{ij,kl}$ as a function of DC source-drain voltage $V_{\mathrm{sd}}$, and magnetic field $B$ up to 6 T. We employed low-frequency lock-in technique with AC excitation voltage $V_\mathrm{AC}=200\;\si{\micro\volt}$, which was reduced by the finite contact resistances to less then $10\;\si{\micro\volt}$ on the sample itself.  One of the current contacts, labelled with ($i,j$), was always connected to virtual ground via the low-noise trans-impedance amplifier \cite{*[{Experimental setup was similar to that described in }] [{ see also }] Kretinin2011,*Kretinin2012}. Conductance $G_{ij,kl}$ is defined here as AC current $I_{ij}$ divided by AC voltage \emph{amplitude} $V_{kl}$, measured on ($k,l$) terminals \footnote{AC voltage amplitude $V_{kl}$ was obtained as $\sqrt {X^2+Y^2}$, where $X$ and $Y$ were real and imaginary parts of measured signal, AC current $I_{ij}$ was driven at reference frequency $f=135.91\,\si {\hertz }$.}. Each collected data point was an average over several read-outs performed at constant magnetic field. The \emph{relative phase} $\theta$ of $V_{\mathrm{AC}}$ and $V_{kl}$ low-frequency signals was also simultaneously recorded. Measurements were carried out in the cryo-free dilution refrigerator Triton 400 at base temperatures $T$ ranging from $0.01$ to $0.3\;\si{\kelvin}$, and performed in the dark with \emph{no illumination}. 

The electron density $n_{\mathrm{2D}}=\num{3.1E11}\:\si{\per\centi\meter\squared}$, which we used for calculations, has been determined from the \emph{quasi-}Hall configuration explained in Fig~\ref{fig:R1628+sample}. Knowing $n_{\mathrm{2D}}$, we have also estimated the carrier mobility $\mu_{\mathrm{H}}=\num{1.9E5}\ \si{\centi\meter\squared\per\ensuremath{(\volt\second)}}$ and electron mean free path $\ell=1.8\;\si{\micro\meter}$ for our sample. Fig~\ref{fig:R1628+sample} shows Hall resistance $R_{\mathrm{H}}$ vs magnetic field $B$, where the quantized plateaus are observed for filling factors $\nu=8,9$ and $10$. The number of occupied LLs and the magnetic field values, for which $R_{\mathrm{H}}$ steps occur, agrees well with the results of calculations, displayed in Figs~\ref{fig:LLcalc} and \ref{fig:edge-1}. However, Hall quantization is not perfect and it entirely disappears for $|B|>2\;\si{\tesla}$.

Figure \ref{fig:R1628+sample} also shows the relative phase $\Theta$ of Hall voltage as a function of magnetic field. In principle, $\Theta$ depends on $R_{\mathrm{H}}$ and $\mathrm{AC}$ frequency $\omega$ via \textit{external} capacitances of a particular experimental setup. For four-terminal Hall measurements the change of phase is given by $\delta\Theta=\tan^{-1}(R_{\mathrm{H}} \omega C_{\mathrm{e}})$, where $C_{\mathrm{e}}$ is the  capacitance of a single connecting cable \cite{Hernandez2014}. In our experimental conditions $C_{\mathrm{e}}\approx 400\ \si{\pico\farad}$, as measured at $300\ \si{\kelvin}$, therefore $\delta\Theta \approx \ang{0.5}$ for $R_{\mathrm{H}}=R_{\mathrm{K}}=h/e^2$. Such a small value is negligible and indeed, $\Theta$ does not change monotonically with $R_{\mathrm{H}}$ but on an average remains equal to $\piup$, as it is expected for $\omega \rightarrow 0$, because Hall voltage is negative for $B<0$\footnote{A small increase of $\Theta$ for $B\rightarrow 0$ is observed because the ($I$-$V$) characteristics of electrical contacts were non-linear for $|B|\lesssim 0.02\ \si{\tesla}$.}. 

\begin{figure}
	\begin{centering}
		\includegraphics[scale=1]{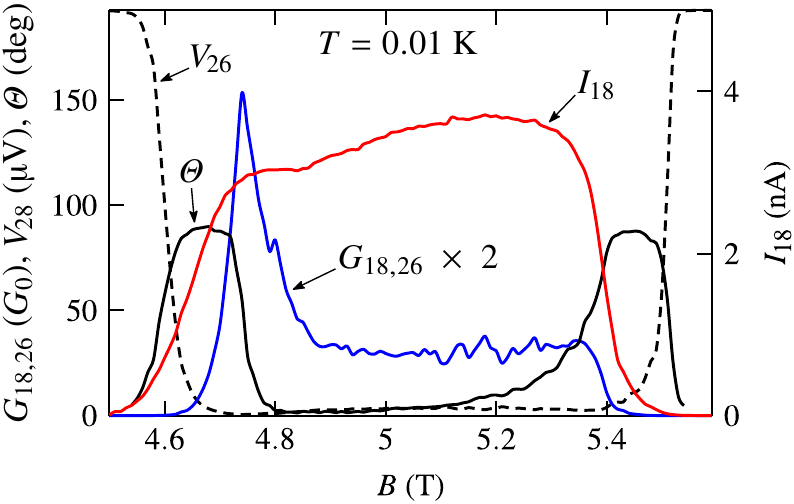} 
		\par\end{centering}
	\caption{Conductance $G_{18,26}$, voltage $V_{26}$, phase $\Theta$ (left axis) and current $I_{18}$ (right axis) vs magnetic field $B$. It is an enlarged part of Fig.~\ref{fig:G1826-G-Theta}, which corresponds to the transition from filing factor $\nu=3$ to $\nu=2$.} \label{fig:G1826-G-Theta-V-I}
\end{figure}

However, phase $\Theta$ is not constant as a function of magnetic field. We observed a rapid fluctuations of phase signal, which amplitude became very large for $|B|>1.5\ \si{\tesla}$. We attribute this effect to the consecutive appearance and disappearance of \textit{internal} capacitances, induced by magnetic field and related to the formation of edge currents in the specific geometry of our microstructure. It is explained below by analyzing conductance, rather than resistance data. 

For analysis, we chose a symmetric, \emph{quasi-}local configuration, in which the wider wires (1) and (8) were used as current contacts and the narrower constrictions (2) and (6) played the role of voltage probes. Figure~\ref{fig:G1826-G-Theta} shows the differential magnetoconductance $G_{18,26}$ obtained at $T=0.01\ \si{\kelvin}$ together with the AC current $I_{18}$, measured simultaneously. For $B<1\;\si{\tesla}$ we observed very well resolved Shubnikov-de~Haas (SdH) oscillations with a characteristic beating pattern caused by giant spin splitting \cite{Kunc2015}. For higher fields, less regular but very pronounced conductance resonances are visible, which are also related to the position of the Fermi level, as explained below. For example, the last resonance  appears when $E_{\mathrm{F}}$  approaches $E_{1}^{\downarrow}$ energy and filling factor $\nu$ changes from $3$ to $2$.

Furthermore, for $B>2\;\si{\tesla}$ we recognize regions where current comes close to zero, for example $I_{18}<\num{1e-12}\;\si{\ampere}$ at $B=4.25\;\si{\tesla}$. At the same time, voltage $V_{26}$ increases, approaching rms amplitude of the external excitation $V_\mathrm{AC}$. Such zero conductance regions are indicated in Fig.~\ref{fig:G1826-G-Theta} with the greyed background. The comparison with Fig.~\ref{fig:LLcalc} reveals, that those broad conductance \textit{anti-resonances} correspond to the situation when Fermi energy is located between Landau levels.

A strong \textit{anti-resonances} of conductance (sometimes called resistance overshoots) were predicted for non-interacting disordered quantum wires \cite{Masek1989} and observed experimentally in GaAs/AlGaAs two-terminal devices \cite{Wrobel1995, *[{ see also }] Wrobel1996} and narrow Hall-bar structures \cite{Kendirlik2013}. They appear, when the central compressible strip of electron liquid in conducting channel is localized by disorder and provides a route for the backscattering. Here, however, the anti-resonances are rather wide and occur when there is no compressible liquid in the central part of the device. Therefore, the filling factor in terminals must differ considerably from its bulk value, to explain the observed current drop. While it is still possible, a large width of the anti-resonances and a very low value of current at the minima rather suggest, that the edge states are at high magnetic fields decoupled from the macroscopic contacts, when $E_{\mathrm{F}}$ enters the energy gap of bulk states.

\begin{figure}
	\begin{centering}
		\includegraphics[scale=1.0]{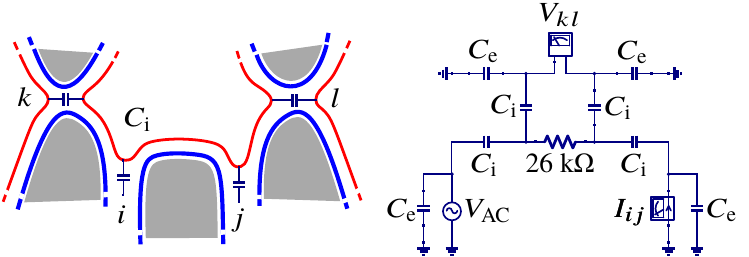} 
		\par\end{centering}
	\caption{Current carrying outer edge channels (blue) and inner edge channels (red), which are disconnected at contact areas but still coupled to them by the internal capacitors $C_{\mathrm{i}}$. The geometry of our sample is reproduced schematically on the left. The simplified electrical equivalent scheme is shown on the right. For cable capacitance $C_{\mathrm{e}}=400\ \si{\pico\farad}$ and $C_{\mathrm{i}}=5\ \si{\pico\farad}$\cite{Hernandez2014} we obtained almost exactly $|\Theta|=\piup/2$.     \label{fig:RC-model}}
\end{figure}

As already mentioned, HgTe/Cr/Au contact pads were evaporated at the boundaries of the sample, however, they were annealed only locally, during indium soldering. Moreover, polycrystals of HgTe do not form a continuous conducting layer, because of their size ($\approx 200\;\si{\nano\meter}$). Therefore, it is very probable that HgTe contacts were well connected to the inner part of the sample but only weakly to the edge. It is known, that such \textit{interior contacts} are connected to the edge currents through Corbino geometry and therefore the coupling is strongest for the \textit{innermost} (i.e. farthest from the edge) channel \cite{Faist1991,Haug1993}. Moreover, quantized resistance plateaus are not observed  for Corbino samples, instead, the conductance drops to zero for integer filling factors \cite{Kobayakawa2013}. Clearly, it is also observed for our device. Therefore we conclude, that HgTe/Cr/Au metal pads form the (unintentionally) interior contacts, which are periodically decoupled from the edge states as a function of magnetic field.

\section{Disconnected edge currents}

While \textit{macroscopic} interior terminals equilibrate rather with the innermost channels, \textit{microscopic} quantum point contacts behave in the opposite way, providing coupling to the outermost (i.e. closest to the edges) currents \cite{Alphenaar1990}. Below, we argue that this difference determines the relative phase $\Theta$ of voltage $V_{26}$ signal, shown in Fig.~\ref{fig:G1826-G-Theta}. The phase does not change much up to $B\approx 1\;\si{\tesla }$, in spite of strong $G$ oscillations, whereas for higher fields it switches between $0$ and $\piup/2$ values. Fig.~\ref{fig:G1826-G-Theta-V-I} provides a detailed comparison of all measured signals for the $\nu=3$ to $\nu=2$ transition region, where chemical potential stays close to the $(1\!\uparrow)$ Landau level.

For $B \approx 4.75\;\si{\tesla}$ we observe the conductance maximum, which is even stronger at $T=0.1\;\si{\kelvin}$, when $G_{18,26}$ reaches extremely high value of $\approx 3000\;G_0$. Vanishing resistance shows that at resonance the outermost edge channel is strongly coupled with both macroscopic and microscopic contacts. In such circumstances, the equilibration between terminals (2) and (6) occurs ($V_{26}) \rightarrow 0$) and the strong peak of conductance is indeed observed ($G_{18,26}\rightarrow\infty$). For higher fields, Fermi level stays close to $(1\!\uparrow)$ LL and sample interior becomes divided into magnetic and non-magnetic domains. The irregular peaks of conductance, observed for still higher fields, are most probably related to the interference pattern of electron waves travelling along a network formed by magnetic domain walls. That picture is supported by the hysteretic behavior of these conductance fluctuations.

For lower fields ($B<4.75\;\si{\tesla}$) current $I_{18}$ starts to decrease, because coupling of the outermost edge channels with the interior contacts is weakened. At the same time, the inner edge channel,  which is still coupled with macroscopic terminals, does not enter the narrower constrictions, as shown schematically in Fig.~\ref{fig:RC-model}. Therefore, the inner edge channel is disconnected by microscopic contacts, however, it is still capacitively coupled with the central part of the device by geometrical and/or electrochemical capacitance $C_{\mathrm{i}}$ \cite{Christen1996}. Such internal coupling schemes have been investigated experimentally in the kHz \cite{Hartland1995, Hernandez2014} and GHz \cite{Gabelli2006, Hashisaka2012} frequency ranges. In our case, the appearance of  the  internal capacitance leads to a phase shift $\deltaup\Theta = \tan^{-1}(1/R_{\mathrm{K}}\omega C_{\mathrm{i}})$, which for low frequencies approaches the `quantized' value $\piup/2$. This conclusion was confirmed by the analysis of a simplified equivalent circuit shown in Fig.~\ref{fig:RC-model}. We applied Kirchhoff's laws only to the innermost channels (shown in red). The fully transmitted outermost channel (shown in blue) were not included in the equivalent circuit, because they mutually equilibrate and do not contribute to the measured values of $V_{26}>0$. For simplicity our model ignores the inter-channel scattering, which may occur in the current terminals \footnote{The calculations were performed using QUCS Studio circuit simulator \url{http://dd6um.darc.de/QucsStudio/qucsstudio.html}}.  

\begin{figure}
	\begin{centering}
		\includegraphics[scale=0.8]{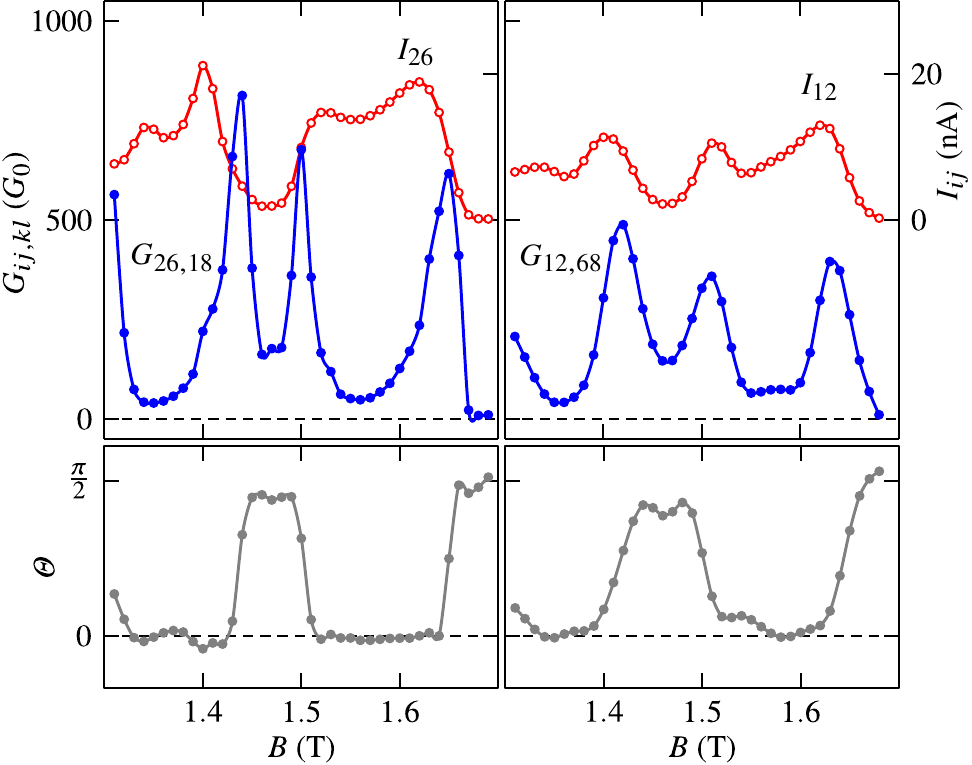} 
		\par\end{centering}
	\caption{Conductance, current and phase vs increasing magnetic field, obtained for $(26,18)$ and $(12,68)$ contact configurations at $T=0.01$~K. $B$ values in the left panel and $\Theta$ values in the right panel were multiplied by $-1$, for clarity. \label{fig:Gijkl-G-Theta-zoom}}
\end{figure}

We apply our analysis also to lower magnetic fields ($B<2\;\si{\tesla}$) where disorder is reduced by screening and edge channels are not completely decoupled from the macroscopic (interior) contacts. If the phase of voltage signal is determined by the position of Fermi energy alone, as explained above, then $\Theta(B)$ pattern will be only weakly dependent on the resistive loads and contact capacitances. It is indeed the case, as it follows from Fig.~\ref{fig:Gijkl-G-Theta-zoom}, were results for ($26,18$) and ($12,68$) configurations are shown. In both cases narrow contact (2) plays a role of the current terminal, therefore $I_{ij}$ fluctuates more strongly and on an average is smaller as compared to corresponding ($18,26$) data. Nevertheless, the main features of $\Theta(B)$ pattern are very similar for all configurations. Phase approaches $\piup/2$ value when $E_F$ is close to ($2$$\uparrow$) Landau level, as if follows from Figs~\ref{fig:LLcalc} and \ref{fig:edge-1}. 

Interestingly,  $\piup/2$ phase shift of even lower frequency ($10.7\ \si{\hertz}$) signal have been observed during the measurements of longitudinal resistance in the QHE regime at filling factor $\nu=1$ for a standard macroscopic Hall-bar device \cite{Desrat2000}. Authors attributed the existence of a zero frequency phase shift to a predominantly inductive load \cite{*[{See also discussion in }] [{}] Melcher2001}. For simplicity, our model ignores the inductance like contributions to admittance \cite{Christen1996}. In any case, however, a small phase shift ($\Theta \approx 0$) guarantees that edge channels are resistively coupled to contact areas.

\begin{figure}
	\begin{centering}
		\includegraphics[scale=0.8]{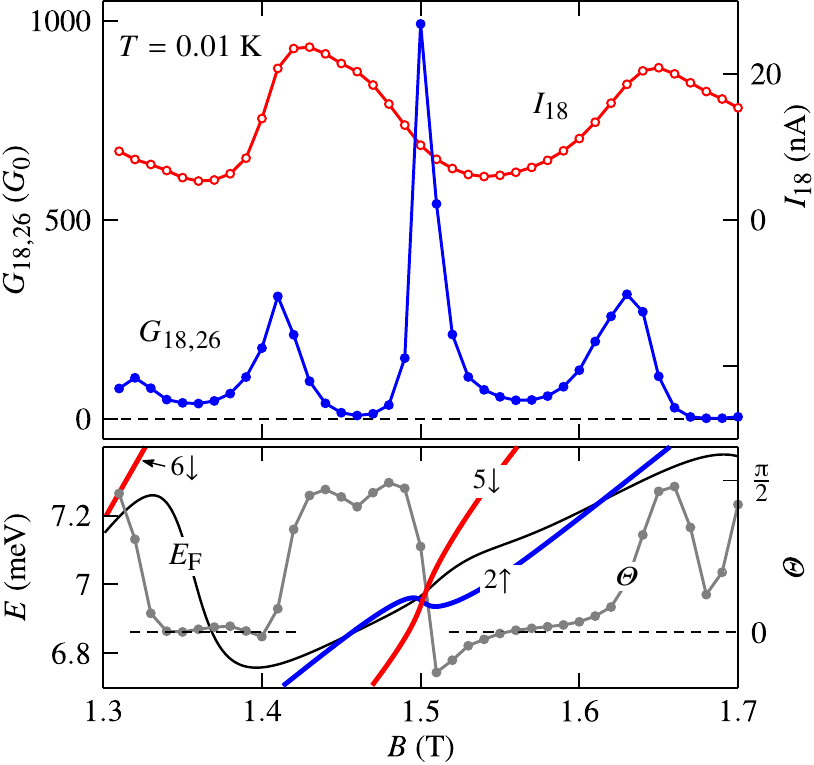} 
		\par\end{centering}
	\caption{Detailed view of the magnetoconductance resonances observed around $B=1.5\ \si{\tesla}$. Current $I_{18}$ data are also shown (as indicated in the upper panel). Phase $\Theta$ vs $B$ is presented below (right axis) together with the bulk energies of the $E_{5}^{\downarrow}$, $E_{2}^{\uparrow}$ and $E_{6}^{\downarrow}$ Landau levels, along with the position of the Fermi energy $E_{\mathrm{F}}$ (left axis). \label{fig:G1826-G-Theta-zoom}}
\end{figure}

\section{Local QHFM transition}

In the following, we concentrate on the local QHFM transition, which corresponds to the pronounced conductance maximum of $G_{18,26}\approx1000$ (in $G_{0}=2e^{2}/h$ units) observed at $B=1.5\ \si{\tesla}$. Figure~\ref{fig:G1826-G-Theta-zoom} shows an enlarged view of three adjacent conductance resonances, occurring at $B=1.41\thinspace,1.50\;\mathrm{and}\;1.63\;\si{\tesla}$. 
We attribute central peak to the crossing of the $E_{5}^{\downarrow}$ and $E_{2}^{\uparrow}$ Landau levels, as shown in the energy diagram of the lower panel. For this, the relative energies of spin-down and spin-up states together with the position of the Fermi level $E_{\mathrm{F}}$ were calculated self-consistently with Mn content $x$ taken as the only fitting parameter. We obtained $x=0.00964$, which is in reasonable agreement with other features observed for the whole range of magnetic fields, as discussed above. The lower panel shows also the relative phase $\Theta$ of voltage signal vs magnetic field $B$ for this configuration. According to the previous discussion, $\Theta \approx 0$ when Fermi energy is located between Landau levels and approaches $\piup/2$ value when $E_{\mathrm{F}}$ is close to one of the quantized energy states.  We would like to stress, that for the central conductance peak $\Theta\approx 0$ and then changes abruptly when Landau levels  $E_{5}^{\downarrow}$ and $E_{2}^{\uparrow}$ alter their positions. At the same time, current $I_{18}$ (i.e. load resistance) practically does not change over the resonance width. 
			
\begin{figure}
	\begin{centering}
		\includegraphics[scale=0.8]{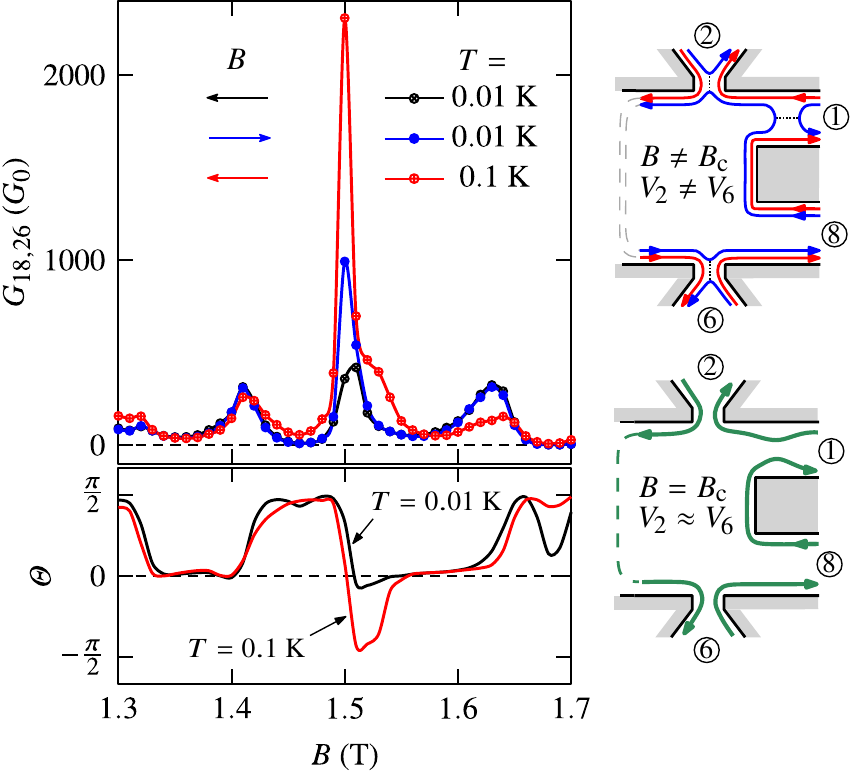} 
		\par\end{centering}
	\caption{(left) $G_{18,26}$ resonances observed around $B=1.5\ \si{\tesla}$ for two temperatures and for opposite directions of magnetic field change (as indicated). The phase $\Theta$ vs decreasing magnetic field is shown below. (right) A schematic view of the \emph{outermost} edge currents at the resonance ($B=B_{\mathrm{c}}$) and off-resonance ($B\protect\neq B_{\mathrm{c}}$). Inter-channel scattering is marked with dotted lines. Edge states, which flow closer to the boundaries, are not shown. \label{fig:G1826-G-peak}}
\end{figure}

The properties of the central conductance peak, observed for local ($18,26$) configuration, are further examined in Fig~\ref{fig:G1826-G-peak}. The middle resonance is higher and narrower than its neighbors. Moreover, it shows hysteretic behavior together with a strong and unusual temperature dependence. At  $T=0.1\;\si{\kelvin}$ conductance increases and reaches $G\approx 2300\;G_0$ value. Note also, that at resonance we still have $\Theta \approx 0$, however, the off-resonance phase changes more abruptly towards $-\piup/2$ value as compared to data obtained at $T=0.01\;\si{\kelvin}$.
		
We deduce that the shape of resonance at $B_{\mathrm{c}}=1.50\;\si{\tesla}$, which is much narrower than neighboring maxima, is a direct consequence of degenerate edge channel reconstruction. This is explained in the right part of Fig.~\ref{fig:G1826-G-peak}, where the upper drawing shows a schematic of our device for $B\lesssim B_{\mathrm{c}}$. Near the resonance, non-interacting edge currents should lie very close to each other. However, Coulomb repulsion pushes the state $E_{2}^{\uparrow}$ away from the edge (and away from state $E_{5}^{\downarrow}\;$) by a distance of approximately $\Delta^{*}$. Consequently, there is a non-zero probability that outermost channel emanated from contact (8) is back-scattered at the entrance contact (1), which is narrower. As a result, equilibration among edge currents which were initially in balance with different current contacts occurs. Due to scattering, which is marked with dotted lines in the figure, probe (2) in not in equilibrium with contact (1) and there is a considerable phase shift between the voltage $V_{26}$ and current $I_{18}\;$ signals. The same picture applies equally well for $B\gtrsim B_{\mathrm{c}}$, the only difference being that $\Delta^{*}$ abruptly changes sign. This explains why conductance resonances related to level crossings are very narrow. To understand the situation at resonance, suppose that at $B=B_{\mathrm{c}}$ the edge currents can overlap ($\Delta^{*}\approx0$) as illustrated schematically in the the lower-right part of Fig.~\ref{fig:G1826-G-peak}. Then, the degenerate channel is closer to the edge and back-scattering caused by differences of width and by disorder is considerably weaker. In the ideal situation, probe (2) is in equilibrium with contact (1) and probe (6) is balanced with probe (2). Therefore, $V_{2}\approx V_{6}$ and differential conductance $G_{18,26}$ is very large with a small current-voltage phase shift $\Theta$. This picture is in agreement with experimental results and explains the large amplitude of resonances associated with the local QHFM transitions.

As already mentioned, at $T=0.1\;\si{\kelvin}$ the conductance peak is higher than at lower temperatures, though it is difficult to find systematic trends because of hysteretic behavior. In particular, some maxima are different whether magnetic field is raised or lowered as seen in Fig.~\ref{fig:G1826-G-peak}. For sharp resonances, this is apparently related to the presence of a narrow hysteresis loop, similar to that shown in Fig.~\ref{fig:edge-1}c with $V_{\mathrm{sd}}$ replaced by $B$. Freire and Egues \cite{Freire2007} predicted such hysteresis for Cd(Mn)Te materials when electron density is globally tuned to the QHFM transition. In our opinion such a mechanism can also operate locally for any carrier concentration, since LLs coincide with the chemical potential at the sample edge. Therefore, the observed hysteresis may be related to the a metastable channel reconstruction, but also to the pinning of 1D magnetic domains which are formed when $\Delta^{*}$ changes sign along the boundaries. Such effects may be a source of non-linearity in quantum transport and indeed, the Onsager relations are fulfilled only approximately for conductance resonances, see Fig.~\ref{fig:Gijkl-G-Theta-zoom}.

\section{Crossing of edge currents}

\begin{figure}
	\begin{centering}
		\includegraphics[scale=0.8]{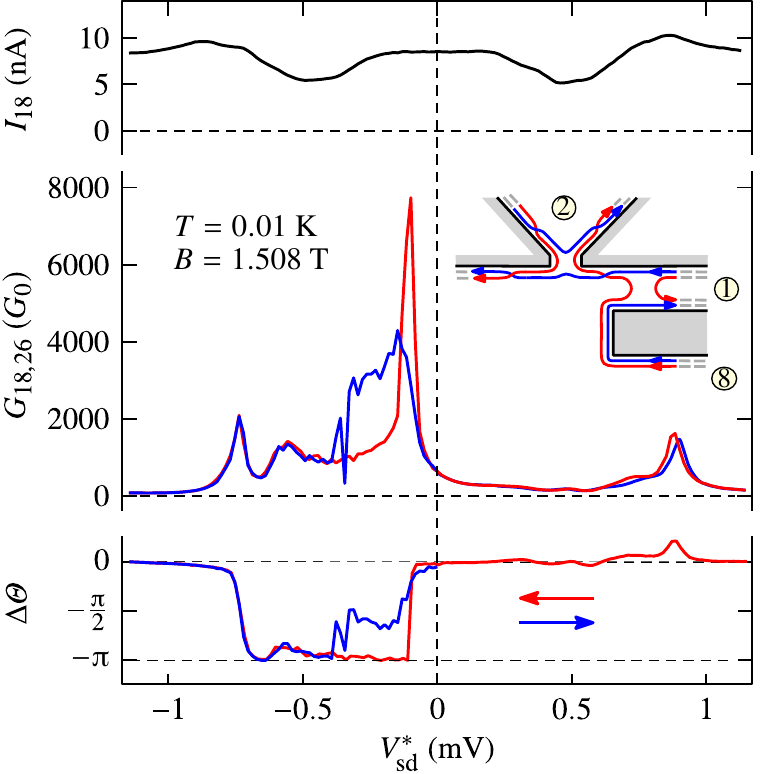} 
		\par\end{centering}
	\caption{$G_{18,26}$ and $I_{18}$ vs source-drain voltage $V_{\mathrm{sd}}^{*}$ recorded at constant magnetic field tuned slightly off-resonance. The relative change of the phase of the AC voltage signal is shown below. Arrows indicate the opposite directions of the $V_{\mathrm{sd}}^{*}$ sweep. The inset shows a schematic of edge currents at local crossings which results in $\Deltaup\Theta=\pm\piup$. Channel ordering at the entrance to the voltage probe corresponds to Fig.~\ref{fig:edge-1}a. \label{fig:G1826-Vsd-10mK}}
\end{figure}

Rijkels and Bauer \cite{Rijkels1994} suggested that such a change of sign can be created on demand when spatially separated edge currents are brought \emph{out of equilibrium}. Then, the edge states cross at certain points giving rise to the formation of magnetic domain walls. Those authors also pointed out that a chemical potential difference can be most easily created by the introduction of a spin selective barrier across one of the current contacts. Such circumstances are naturally realized in our sample, since the wire contact (1) is about $1\;\si{\micro\meter}$ narrower than the second current terminal (8) -- see the top-right schematic in Fig.~\ref{fig:G1826-G-peak}. Therefore, a non-equilibrium population of outermost edge currents was accomplished by applying a source-drain voltage difference on contacts (1) and (8).

The experiment was conducted as follows. We again focused on the conductance maximum at $B_{\mathrm{c}}=1.50\;\si{\tesla}$ and the magnetic field value was fixed at $B=1.508\;\si{\tesla}$ in order to slightly \emph{detune} from the resonance peak. This corresponds to the situation shown schematically in Fig.~\ref{fig:edge-1}b for $B\gtrsim B_{\mathrm{c}}$, where the separation distance of edge currents is $\Delta^{*}<0$. By applying a source-drain voltage one hopes to reduce the absolute value of separation, \emph{tune back} to the resonance and eventually force the edge channels to cross. See Fig.~\ref{fig:edge-1}c.

The results are summarized in Fig.~\ref{fig:G1826-Vsd-10mK}, where upper panel shows $G_{18,26}$ as a function of the chemical potential difference between contacts (1) and (8), denoted as $V_{\mathrm{sd}}^{*}$. Let us focus on the red line. For negative values of $V_{\mathrm{sd}}^{*}$, we observe a very high and narrow conductance peak at $V_{\mathrm{sd}}^{*}=-0.1\;\si{\milli\volt}$ having a half-width of $\Deltaup E^{*}\approx60\;\si{\micro\electronvolt}$. It is very similar to the resonances obtained as a function of magnetic field. As explained before, a maximal conductance corresponds to a minimal distance between edge currents. Therefore, further increase of $|V_{\mathrm{sd}}^{*}|$ should induce channel swapping and a decrease of conductance.

\begin{figure}
	\begin{centering}
		\includegraphics[scale=0.75]{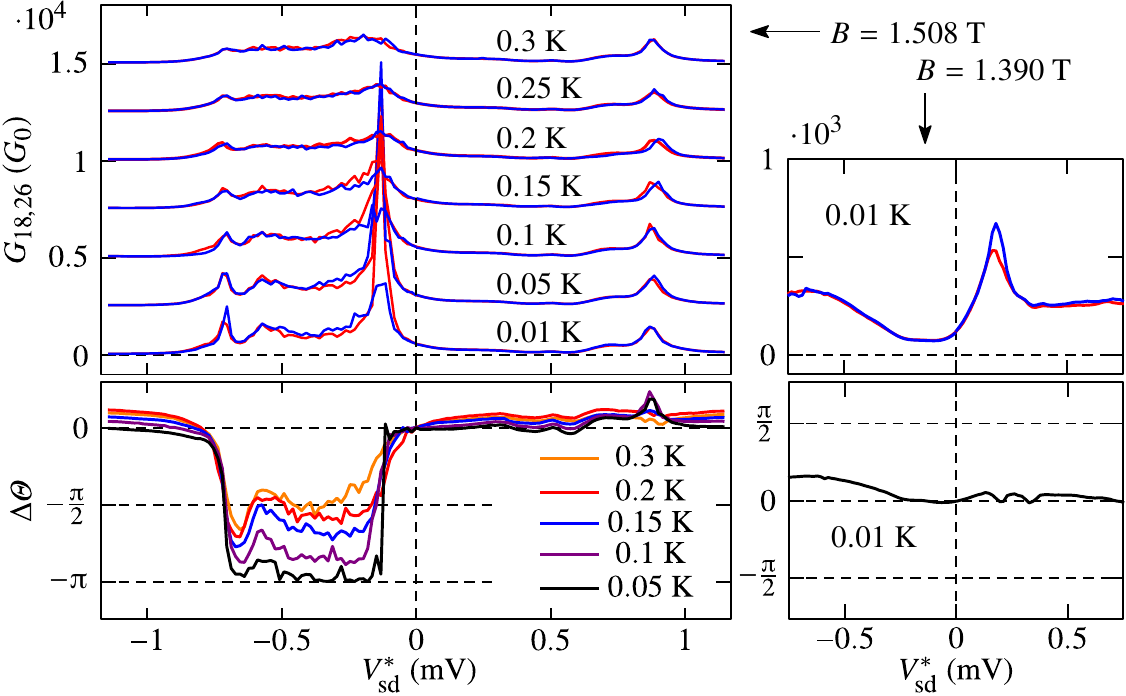} 
		\par\end{centering}
	\caption{Conductance $G_{18,26}$ (upper panel) and relative change of voltage phase (lower panel) vs source-drain voltage $V_{\mathrm{sd}}^{*}$ recorded at magnetic field $B=1.508\;\si{\tesla}$ for different temperatures. In the upper panel, data are shown for both directions of the $V_{\mathrm{sd}}^{*}$ sweep, and were shifted vertically for clarity. The $\Deltaup\Theta$ data are shown for decreasing source-drain voltage only. \label{fig:G1826-Vsd-T}}
\end{figure}

This is indeed what was observed in the experiment. The lower panel shows $\Deltaup\Theta$, the change of the AC voltage phase $\Theta$ relative to its value at $V_{\mathrm{sd}}^{*}=0$. This phase changes abruptly by 180 degrees just after passing the resonance, as would be expected when edge currents cross at some point and interchange their positions. In our case, the voltage probe (2) is in equilibrium with the current contact (8) when the intersection is created and the phase of the $V_{26}$ signal is shifted by $\pm\piup$. See the inset of Fig.~\ref{fig:G1826-Vsd-10mK}, keeping in mind that not only one (as shown) but also an \textit{odd} number of crossings located between probes, will lead to $|\Deltaup\Theta|=\piup$. Interestingly, $\Deltaup\Theta$ is close to $-\piup$ down to $V_{\mathrm{sd}}^{*}\approx-0.7\;\si{\milli\volt}$, suggesting that  channels are still connected to reservoirs. Note also, that ($I$-$V_{\mathrm{sd}}^{*}$) curve is almost exactly symmetrical around zero as it is evident from $I_{18}$ data presented in the upper panel. Therefore, the abrupt phase change must be attributed to channel reordering.
	
As pointed out by Rijkels and Bauer, each crossing point causes the topological problem of how to connect edge channels to the macroscopic terminals, which in equilibrium emanate states with a normal ordering of spins. Therefore, the channels have to switch again somewhere, and one of such possible crossings is shown in the inset on the left of probe (2), two others on the other side. The number of topological defects depends on mutual electrostatic repulsion  \cite{Rijkels1994} and our data suggest that the constriction itself (which plays a role of voltage probe) may induce a channel switch. In that case, channel ordering is reversed in the vicinity of a smooth saddle-point potential, as compared to the outer regions of a constriction. Quite recently it was suggested that for low Zeeman energies the ordering of the edge modes switches abruptly at the narrowing, while the arrangement in wider areas remains intact \cite{Khanna2017}. Such mechanism, driven by spin exchange interaction may operate here.

Figure \ref{fig:G1826-Vsd-10mK} also shows results obtained when the source-drain voltage $V_{\mathrm{sd}}^{*}$ was increased from negative to positive values. For $V_{\mathrm{sd}}^{*}\gtrsim-0.4\;\si{\milli\volt}$ conductance reveals strong instabilities and $\Deltaup\Theta$ deviates from $-\piup$. It appears that the system was trapped in a metastable state for which channel coupling to reservoirs is different. However, the hysteretic behavior shown in Fig.~\ref{fig:G1826-Vsd-10mK} represents a rather extreme case. Typically, conductance instabilities are weaker, as illustrated in the left part of Fig~\ref{fig:G1826-Vsd-T} where $G(V_{\mathrm{sd}}^{*})$ and $\Deltaup\Theta(V_{\mathrm{sd}}^{*})$ for $B=1.508\;\si{\tesla}$ data are shown. Analogous data for the neighboring conductance peak (see Fig~\ref{fig:G1826-G-peak}), recorded at $B=1.390\;\si{\tesla}$, are shown to the right. Also here, we can tune back to the resonance with source-drain voltage (this time for $V_{\mathrm{sd}}^{*}>0$), however, the phase of the voltage signal practically does not change when conductance passes through its maximum. This proves that the abrupt change of $\Theta$ by $\piup$ is characteristic for LLs crossings only.

Data for channel crossing at $B=1.508\;\si{\tesla}$ are also shown as a function of temperature. For $T>0.1\;\si{\kelvin}$ the fluctuations of the peak value are smaller and the conductance resonance gradually disappears. However, at $T=0.3\;\si{\kelvin}$ it is still clearly visible. A strong temperature dependence of $\Deltaup\Theta$ is also observed for the range of source-drain voltages that corresponds to a channel crossing. At $T=0.05\;\si{\kelvin}$, the phase changes when detuned off-resonance by almost exactly 180 degrees, but at higher temperatures it is continuously reduced, reaching about 90 degrees seen at $T=0.3\;\si{\kelvin}$. Such temperature averaging of phase (over the extreme values of $0$ and $\piup$) happens when $k_{\mathrm{B}}T$ is of the order of $\Deltaup E^{*}$, the energy difference between spin-up and spin-down electrons at the edge of a sample. For $V_{\mathrm{sd}}^{*}>0$, the temperature dependence of $G$ and $\Deltaup\Theta$ is much weaker, which is expected since $\Delta^{*}$ increases and channel crossing does not occur.

All this fits very well to the picture of edge current reconstruction and local channel crossing predicted by theory. There are, however, some noticeable differences. First, the almost symmetrical hysteresis shown schematically in Fig~\ref{fig:edge-1}c occurs only for negative values of source-drain voltage and therefore in experiment is highly asymmetric. Second, the bistability predicted by Rijkels and Bauer does not manifest itself clearly, because abrupt changes of $\Deltaup\Theta$ from $-\piup$ to 0 are not observed. Finally, $\Deltaup E^{*}$ is much smaller than calculated by theory \cite{Dempsey1993,Rijkels1994}. However, it is known that the splitting is overestimated in the Hartree-Fock approximation and in our case spin levels belong to different orbitals, therefore exchange effects are strongly reduced. On the other hand, the anticrossing gap between the edge channels may be enhanced by spin-orbit coupling as suggested previously for bulk states in \cite{Kazakov2016}. There, $\Delta_{\mathrm{SO}}=70\;\si{\micro\electronvolt}$ was obtained for the $E_{0}^{\uparrow}$ and $E_{1}^{\downarrow}$ Landau levels.

In fact, our experiment covers a broader range of phenomena than have been considered by the theories discussed above. In the non-linear transport regime we observed two additional conductance maxima, on both sides of the central resonance peak, as clearly visible in Figs \ref{fig:G1826-Vsd-10mK} and \ref{fig:G1826-Vsd-T}. We attribute them to the crossings of Landau levels, which are adjacent in energy and gradually appear in a conductance window when source-drain voltage is applied. The left resonance appears to be related to an intersection of the states $E_{4}^{\downarrow}$ and $E_{1}^{\uparrow}$, located below the Fermi level in equilibrium. Apart from a smaller amplitude this peak is very similar to the central resonance with regard to temperature dependence, fluctuations of the maximal value. Most importantly, it is also connected with a change of the AC voltage signal phase by 180 degrees. The right resonance is most probably related to the intersection of the $E_{6}^{\downarrow}$ and $E_{3}^{\uparrow}$ states, which are not occupied for $V_{\mathrm{sd}}^{*}=0$. In this case, the observed phase changes and temperature dependence of conductance are much weaker.

\section{Summary}

In summary, we have presented the results of low temperature magneto-transport measurements, obtained using a low-frequency $AC$ lock-in technique. Measurements were carried out on a \textit{quasi-}ballistic micro-structure created on a very good quality $n-$type Cd$_{1-x}$Mn$_{x}$Te diluted magnetic quantum well. We analysed not only amplitude of conductance $G_{ij,kl}$, but also phase $\Theta$ of the measured signal. We have observed sudden jumps of phase between $0$ and $\piup/2$ values, which we explained as a consecutive coupling and decoupling of the inner edge channels to the current and voltage contact areas. Most importantly, for $B=1.5\ \si{\tesla}$ (and $\Theta \approx 0$) we have observed high and narrow conductance resonance attributed to transition to the QHFM phase, which occurs at sample boundary and is accompanied by the \emph{reconstruction} of spin-degenerate edge states. Furthermore, we show that a \emph{local crossing or reordering} of spatially separated chiral channels,  which is associated with the change of $|\Theta|$ by $\piup$, can be induced on demand by applying a DC source-drain voltage.

\begin{acknowledgments}
The research was partially supported by the National Science Centre (Poland) through grants No. DEC- 2012/06/A/ST3/00247, 2016/23/N/ST3/03500 and by the Foundation for Polish Science through the IRA Programme co-financed by the EU within SG OP. Authors wish to thank P. Deuar and T. Dietl for valuable remarks.
\end{acknowledgments}

\end{document}